\documentstyle[preprint,aps,amssymb,amsfonts,prbbib,psfig]{revtex}
\tightenlines
\begin{document}
\draft
\title{Solution of Poisson's equation for finite systems using plane wave
methods}
\author{Alberto Castro and Angel Rubio}
\address{Departamento de F\'{\i}sica Te\'{o}rica, Universidad de Valladolid,
E-47011 Valladolid, Spain}
\author{M. J. Stott}
\address{Physics Department, Queen's University, Kingston, Ontario, Canada
K7L 3N6}
\date{\today}
\maketitle

\begin{abstract}
Reciprocal space methods for solving Poisson's equation 
for finite charge distributions are 
investigated. Improvements to previous proposals are presented, and 
their performance is compared in the context of a real-space 
density functional theory code. Two basic methodologies are followed:
calculation of correction terms, and imposition of a cut-off to the
Coulomb potential. We conclude that these methods can 
be safely applied to finite or aperiodic systems with a reasonable control
of speed and accuracy.
\end{abstract}


\section{Introduction}

Density-functional theory \cite{Hohenberg_Kohn_64,Kohn_Sham_65}
in its time-dependent \cite{Runge_Gross_84} as well as ground 
or time-independent forms has proved to be an efficient method for 
treating electron-electron interactions and has been applied successfully to 
finite systems such as clusters, \cite{Rubio_etal_97} 
to bulk systems or surfaces, 
and to aperiodic systems such as defects. \cite{Brack_93}
However, the high computational 
cost of treating large systems places a practical limit on the size of 
systems that can be studied. 

The use of pseudopotentials \cite{Pickett_89,Cohen_82}
enhances the performance of this sort of
calculation by avoiding an explicit treatment of the Kohn-Sham orbitals 
associated with the core. Furthermore, the smoothness of the resulting valence 
pseudowavefunctions allows the use of a plane wave basis for describing 
them, and consequently also the electronic density. A plane wave basis is
particularly attractive because it allows use of the fast fourier transform
 (FFT), for rapid and memory efficient transformations.

A discrete but truncated set of plane waves based on the reciprocal lattice 
is one natural basis set for a periodic system. However, for finite systems, 
or more generally, for systems lacking periodicity such as defects in 
solids, the use of a discrete set of plane waves will generate periodic 
images of the finite cell to be studied. In the case of a finite system this
leads to a problem in the calculation of the electrostatic potential due to the 
electrons, the so-called Hartree potential, due to 
the long range of the Coulomb interaction. Nevertheless, discrete plane 
wave basis sets are often used for finite systems because of the great 
efficiency of the FFT, and errors in the Hartree potential due to periodic 
images are usually ignored, or reduced by increasing the size of the 
supercell. These spurious effects might seriously affect the calculated
equilibrium structure and dynamics of weakly bounded molecules or 
clusters, eg. water. 
However, several methods have been proposed recently for treating 
this problem. \cite{Lauritsch_Reinhard_94,Onida_etal_95,Makov_Payne_95,%
Jarvis_etal_97,Shultz_99,Shultz_00}

Our purpose here is to compare four methods for solving Poisson's equation 
for finite systems. One of them is an iterative, real-space method based on 
finite differences and conjugate-gradients minimization, which obviously 
doesn't suffer from the problems related to periodic images. The other three 
use discrete plane wave basis sets and FFT's, but treat the cell-to-cell 
interactions in different ways. Two of these plane wave methods impose 
a cut-off to the Coulomb interaction in real space, and have been 
described elsewhere. \cite{Jarvis_etal_97}
However, for one of these, which uses what
we term a cubic cut-off, we have found significant improvements 
which we believe 
will be of interest to practitioners. The third plane wave method 
has been developed and tested by us, but we have found a close relationship 
between it and the local moment countercharge (LMCC) method proposed 
by Schultz, \cite{Shultz_99}
and also to the Fourier analysis with long range 
forces (FALR) method of Lauritsch and Reinhard.
\cite{Lauritsch_Reinhard_94} However, our 
scheme is formulated more generally, and allows for better control of errors.

In order to compare the performance of the different methods we have studied 
some exactly soluble model systems, and $\text{NaCl}$ and 
$\text{Na}_{10}^{+2}$ molecules which, because of their polar or charged 
nature, are difficult to treat with plane wave 
methods. Of particular interest is the way the computational time scales 
with system size. Although all the plane wave methods scale as a few times 
$N\log N$, where $N$ is the number of real-space mesh points, the 
proportionality factor varies substantially from method to method. We 
shall compare the speed and memory requirements of the methods, and how 
these scale with the size of the systems.

The plane wave schemes we discuss are intended mainly to deal with neutral 
or charged molecules or clusters in free space and could be implemented 
directly in existing {\em ab initio} plane wave or real-space 
codes, but they could also 
be used in those LCAO basis set codes which base the calculation of the 
Hartree potential on the FFT. General subroutines for calculating the 
Hartree potential using these methods are available upon request from the 
authors or can be downloaded from the web page.\cite{webpage}

Short theoretical descriptions of the plane wave methods are given in 
section II where we emphasize the improvements we have developed. Section 
III presents and compares the results for the Hartree potential calculated 
using the different methods, and concluding remarks are made in the final 
section. Atomic units are used unless otherwise stated.

\section{Theoretical description of plane-wave methods}

\subsection{Uncorrected calculations}

The solution of Poisson's equation, $\nabla^2 V_H + 4\pi n=0$, which goes 
to zero at infinity, for a charge density $n$, localized within a cell $C$ of 
volume $\Omega$, is given by 

\begin{equation}\label{eq:v_hartree}
V_H[n,{\bf r}] = \int d{\bf r}' \frac{n({\bf r}')}{|{\bf r}'-{\bf r}|}.
\end{equation}

Within the cell $n$ and $V_H$ may be expressed as Fourier series:
$\displaystyle n({\bf r})=\Omega^{-1} \sum_{{\bf G}} e^{i{\bf G}{\bf r}}
\tilde{n}({\bf G})$
where $\displaystyle \tilde{n}({\bf G}) 
= \int d{\bf r}n({\bf r})e^{-i{\bf Gr}}$, 
and similarly for $V_H$, and where the ${\bf G}$ vectors are reciprocal 
vectors of the lattice formed by repeating the cell $C$. If the Fourier 
coefficients, $\tilde{n}({\bf G})$, are negligible for $G$ larger than some 
cut-off so that the sums over ${\bf G}$ may be truncated, then the 
$\tilde{n}({\bf G})$ and the $n({\bf r})$ points are related through a discrete 
Fourier transform. This amounts to approximating the integral over the 
cell in the definition of $\tilde{n}({\bf G})$ 
by the trapezium rule, a point to 
which we shall return later. However, $n({\bf r})$ given by the Fourier sum 
is periodic so that the straightforward substitution into
Eq.\ (\ref{eq:v_hartree}) gives a potential:

\begin{equation}\label{eq:conv}
V[n,{\bf r}] = \frac{4\pi}{\Omega} \sum_{{\bf G}\ne {\bf 0}} 
\frac{\tilde{n}({\bf G})}{{\bf G}^2} e^{i{\bf G}{\bf r}},
\end{equation}
which differs from $V_H$. But the merit of V is that it can be calculated 
using the the very efficient FFT with its $NlogN$ scaling, and so we now 
modify or apply corrections to Eq.\ (\ref{eq:conv}) so that it can be used to 
obtain $V_H$. Two aspects of $V$ given by Eq.\ (\ref{eq:conv}) require 
attention.

\begin{itemize}

\item The ${\bf G}={\bf 0}$ component in Eq.\ (\ref{eq:conv}) is 
arbitrarily set to zero. For a charged system, this corresponds
physically to introducing a uniform compensating charge background, $b$,
so that 
the system is electrically neutral. For a neutral system, this means 
that the boundary condition, $V({\bf r} \to \infty)=0$, is not satisfied 
(which, of course, also happens in the charged case).

\item $V$ is the potential due to the charge distribution $n$ in the 
central cell plus that due to the images of $n-b$ in all other cells.

\end{itemize}

Since we are dealing with the electron charge distribution, there is 
obviously a net charge. The fact that the whole system ({\em cores} + 
electrons) may or may not be charged, is irrelevant for the discussion 
presented in this paper. However, it might be important if the 
calculations are total-energy supercell calculations. In this case, 
the system of ion cores is also treated using reciprocal space, so that 
a background of opposite sign has to be added. If the finite 
system is neutral, the effect of the backgrounds cancels. 
The spurious effect of 
higher multipoles, however, will remain. The 
distinction is important, though, because the uniform background
introduces an important error in the total energy
of order $O(L^{-1})$, $L$ being the size of the cell,
whereas the leading effect of 
the presence of the multipoles is the dipole-dipole term,
behaving like $O(L^{-3})$. This is 
shown in the calculations presented in next section.

\subsection{Multipoles-corrections method}

We can start to deal with the cell-to-cell interaction,
by eliminating the effects of net charge. This can be done by subtracting 
from the original charge distribution, $n$, an auxiliary charge distribution 
$n_{aux}$, so that no net charge remains. The potential $V_H$
then becomes:

\begin{equation}
V_H[n]=V_H[n-n_{aux}]+V_H[n_{aux}].
\end{equation}

The term $V_H[n-n_{aux}]$ can be treated using the FFT techniques, and 
then the correction $V_H[n_{aux}]$, calculated explicitly in real 
space, added on. This method is especially convenient if the Fourier 
components of $n_{aux}$ can be calculated analytically, so that the 
expression becomes:

\begin{equation}
V_H[n] \approx V[n] - \psi,
\end{equation}
where $V[n]$ follows the definition given in Eq.\ (\ref{eq:conv}), 
and $\psi=V[n_{aux}]-V_H[n_{aux}]$ is a function which can be calculated 
analytically. The sign $\approx$ denotes that the effect of
higher multipoles is still included. The choice of $n_{aux}$ is 
arbitrary; it could be a uniform density,
or a gaussian centered on the origin, in both cases the function 
$\psi$ can be calculated analytically.\cite{Hummer_96}

This procedure can be easily generalized, to account for the 
interaction of higher multipoles. We merely 
need to add an auxiliary charge distribution which mimics the multipoles 
whose effect we want to subtract. This procedure is called by 
Shultz \cite{Shultz_99} local moment countercharge method (LMCC). Shultz 
accounts for the monopole and dipole corrections through a superposition
of localized Gaussian charge distributions constructed to have the 
same net charge and dipole moment. Higher multipoles can similarly be 
accounted for
by the superposition of additional Gaussian distributions,
but the procedure becomes complicated. A straightforward 
approach which is more easily generalized introduces an auxiliary charge 
distribution in the form:

\begin{equation}\label{eq:def_of_naux}
n_{aux}({\bf r}) = \sum_{l=0}^{\infty} \sum_{m=-l}^{l} n_{lm}({\bf r}),
\end{equation}
where 
\begin{equation}
n_{lm}({\bf r}) = M_{lm} \frac{2^{l+2}}
{a^{2l+3}\sqrt\pi(2l+1)!!}r^le^{-(r/a)^2}Y_{lm}({\bf r}),
\end{equation}
and $M_{lm}$ is the multipole moment of $n$ given by
\begin{equation}
M_{lm} = \int d{\bf r}n({\bf r})r^lY_{lm}({\bf r}).
\end{equation}

The width parameter $a$ is to be chosen so that $n_{aux}$ is negligible 
at the cell boundary. If high order moments are required $a$ could be 
taken to be $l$-dependent, decreasing somewhat with $l$. Note that the
$l=0$ term in Eq.\ (\ref{eq:def_of_naux}) corrects for the net 
charge as described above.

The auxiliary density is localized within a cell and has the same 
multipole moments as $n$. We can now correct for the presence of the 
periodic images of $n$, and obtain for the required Hartree potential 
in the central cell:

\begin{equation}\label{eq:mc_general_formula}
V_{H}[n,{\bf r}] = V[n,{\bf r}] - \sum_{l=0}^{\infty}\sum_{m=-l}^{l} 
M_{lm}\psi_{lm} + V_0, 
\end{equation}
where $V_0$ is a constant shift yet to be determined, and the functions 
$\psi_{lm}$, which are independent of the charge distribution $n$, are 
given by

\begin{equation}\label{eq:mc_corrections}
\psi_{lm}({\bf r}) = \frac{(4\pi)^2}{\Omega}\frac{i^l}{(2l+1)!!}
\sum_{{\bf G\neq 0}}G^{l-2}e^{-a^2G^2/4}Y_{lm}({\bf G})e^{-i{\bf Gr}} 
- \frac{\sqrt\pi 2^{l+3}}{(2l+1)!!} \frac{1}{r^{l+1}}I_l(r/a)Y_{lm}({\bf r}).
\end{equation}

The first term in $\psi_{lm}$ is the periodic potential due to $n_{lm}$ in 
every cell, and the second term subtracts the effect of $n_{lm}$ in the 
central cell. The function $I_l$ is:
\begin{equation}
I_l(x)=\int_{0}^{x} dt t^{2l} e^{-t^2}.
\end{equation}

The procedure for obtaining $V_H$ is to calculate and store the functions 
$\psi_{lm}$ once and for all for as many of the multipoles as are needed 
to achieve the desired precision, then for the particular charge 
distribution $V$ is calculated using FFTs, the $M_{lm}$ are computed, 
and Eqs.\ (\ref{eq:mc_general_formula}) and (\ref{eq:mc_corrections}) 
used to obtain $V_H$ within the central cell apart from the constant shift.

The shift $V_0$ is chosen so that the boundary condition 
$V_H({\bf r} \to \infty )=0$ is satisfied. This we accomplish by computing 
the average value over the surface of the cell of the corrected but 
unshifted potential $V_H$ obtained by simply putting $V_0=0$ in 
Eq.\ (\ref{eq:mc_general_formula}).
For a cubic cell we have:\cite{note_on_sa_1}
\begin{equation}\label{eq:constant_shift}
\overline{V_H} =\frac{1}{6\Omega^{2/3}}\int_{C} d{\bf s}V_H({\bf r}) 
= \frac{1}{6\Omega^{2/3}}\int_{C} d{\bf s} \int d{\bf r'}
\frac{n({\bf r'})}{|{\bf r} - {\bf r'}|}.
\end{equation}

Since $n$ is zero at the boundary we may interchange the order of 
integration in Eq.\ (\ref{eq:constant_shift}) to give:

\begin{equation}
\overline{V_H} = \int d{\bf r'}n({\bf r'})u({\bf r'}),
\end{equation}
where
\begin{equation}
u({\bf r'}) = 
\frac{1}{6\Omega^{2/3}}\int_{C} d{\bf s}\frac{1}{|{\bf r} - {\bf r'}|}
\end{equation}
is the potential inside the cube due to a unit charge uniformly distributed 
over the cube surface. As such, by Gauss's theorem and symmetry, $u$ is 
constant inside the cell and has the value $\alpha/\Omega^{1/3}$ with 
$\alpha=1.586718$ for a cube. If $n$ integrates to $z$, then we have 
finally:
\begin{equation}
\overline{V_H} = \frac{\alpha z}{\Omega^{1/3}}.
\end{equation}
and $V_0$ should be chosen so that 
$\overline{V_H}$ has this value. The calculation 
of the correct shift to satisfy the boundary condition 
requires the computation 
of the surface average of $V$, but since the Kohn-Sham orbitals are 
unaffected by the addition of a constant to the potential, this only needs 
to be performed at the end of a self-consistent cycle.\cite{note_on_sa_2}

\subsection{Cut-off methods}

We now review two established methods,
\cite{Onida_etal_95,Jarvis_etal_97}
based on imposing a 
cut-off on 
the Coulomb interactions. They are exact, but need a bigger cell, which
is a computational drawback as we shall see.

Let us define a new cell $D$, which includes $C$. This new cell will define 
new coefficients ${\bf G}_D$, which are the reciprocal vectors of the 
lattice formed by repeating $D$. Retrieving 
Eq.\ (\ref{eq:conv}), we realize that the function $V[n]$ can also be 
expressed as:
\begin{equation}
V[n,{\bf r}] = \int d{\bf r'}n^{(p)}({\bf r'})\frac{1}{|{\bf r'}-{\bf r}|},
\end{equation}
where $n^{(p)}$ is the function formed by the sum 
of $n$ and all its periodic
repetitions in the superlattice.

We now introduce a truncated Coulomb potential, with the following properties:

\begin{equation}\label{eq:conditions}
f({\bf r} - {\bf r'}) = \left\{ 
     \begin{array}{rl} 
     \frac{1}{|{\bf r} - {\bf r'}|} &,\;\; \text{for } {\bf r} \text{ and } 
{\bf r'} 
\text{ both belonging to the same image of C.}  \\
     0           &,\;\; \text{for } {\bf r} \text{ and } {\bf r'} 
\text{ belonging to different images of } C.
     \end{array}  \right.
\end{equation}

It is easily seen that:

\begin{equation}
\int d{\bf r'}n^{(p)}({\bf r'})f( {\bf r'}-{\bf r} ) = 
V_H[n,{\bf r}]
\end{equation}
for every ${\bf r} \in C$. And thus we can calculate $V_H$ as 
we did for $V$ in Eq.\ (\ref{eq:conv}):

\begin{equation}
V_H({\bf r}) = \frac{4\pi}{\Omega} \sum_{{\bf G}\ne {\bf 0}} 
\tilde{n}({\bf G}) \tilde{f}({\bf G}) e^{i{\bf G}{\bf r}},
\end{equation}
where $\tilde{f}({\bf G}) = \int d{\bf r}f({\bf r})e^{-i{\bf Gr}}$.

Two choices for $D$ and $f$ have been given. One of them 
uses a spherical shape \cite{Onida_etal_95} 
for the cut-off of the Coulombic interaction, and the 
other a cubic shape,\cite{Jarvis_etal_97}
based on the assumption that the original cell 
$C$ is cubic itself.

\subsubsection{Spherical cut-off method}

Let $L_C$ be the length of the side of $C$.
We will define a larger cubic cell of 
side $L_D=(1+\sqrt{3})L_C$, centered on $C$. For this choice of $D$ we define
next the truncated Coulomb interaction: 

\begin{equation}
f( {\bf r}-{\bf r}' ) = \left\{ 
     \begin{array}{rl} 
     \frac{1}{|{\bf r}-{\bf r}'|} & ,\;\;|{\bf r}-{\bf r}'| < \sqrt{3}L_c \\
     0                            & ,\;\;|{\bf r}-{\bf r}'| > \sqrt{3}L_c. 
     \end{array}  \right.
\end{equation}

Defined in this way, $f$ meets the required conditions 
expressed in Eq.\ (\ref{eq:conditions}),
because any 
two points belonging to $C$ are always closer than $\sqrt{3}L_C$, and any 
two points belonging to different images of $C$ are always farther away than
$\sqrt{3}L_C$.

The Fourier transform of $n$ has to be calculated numerically in the 
larger cell 
$D$; however that of $f$ is easily obtained analytically:

\begin{equation}
{\cal F} \lbrace f \rbrace ({\bf G}) = 
4\pi\frac{1 - \cos(G\sqrt{3}L_C)}{G^2}.
\end{equation}

\subsubsection{Cubic cut-off method}

The former proposal is exact; but a very large cell is needed, which 
increases the time to evaluate the FFTs. Reducing $L_D$ introduces spurious 
interactions and thus spoils the precision of the calculations, but if
extremely precise calculation are not needed, a compromise could be reached.

Our aim now is to reduce $L_D$ but maintain an accurate evaluation of $V_H$.
We take the larger cell to have $L_D=2L_C$ and 
the {\em cut-off} Coulomb interaction to be:

\begin{equation}
f( {\bf r}-{\bf r}' ) = \left\{
     \begin{array}{rl}
     \frac{1}{ |{\bf r}-{\bf r}'| } & ,\;\;{\bf r} - {\bf r}' \in D \\ 
     0           & ,\;\;\mbox{otherwise.} 
     \end{array}  \right.
\end{equation}

If ${\bf r}$ and ${\bf r'}$ belong to $C$, ${\bf r}-{\bf r'}$ belongs to $D$.
And if ${\bf r}$ and ${\bf r'}$ belong to different images of $C$, then 
${\bf r}-{\bf r'}$ will not belong to $D$. Thus again $f$ is correctly defined.

The Fourier transform of this function 
$f$ has to 
be calculated numerically, and we face here two drawbacks:
the function has a singularity at the origin, and
is not analytic at the boundary.


Jarvis {\em et al} \cite{Jarvis_etal_97} dealt with the first problem
by integrating and averaging the singularity over a grid unit,
which may not be adequate. The second problem appears to have been
overlooked. In any event, their treatment of the Mg atom using the
cubic cut-off method converges poorly compared with the results with
the spherical cut-off, and they declare a preference for the spherical
cut-off despite the larger cell size required.
However, we shall show how to overcome these difficulties 
so that the cubic cutoff method, to be preferred because of its smaller cell 
size, can be used with great precision.

The singularity at ${\bf r}={\bf r'}$ can be treated as follows:

\begin{equation}\label{eq:cubic_cutoff_int}
\int_{D} d{\bf r} \frac{1}{r}e^{-i{\bf Gr}} = 
\int_{D} d{\bf r} \frac{\text{erf}(r/a)}{r}e^{-i{\bf Gr}} +
\int_{D} d{\bf r} 
\frac{1-\text{erf}(r/a)}{r}e^{-i{\bf Gr}}, 
\end{equation}
where $a$ is chosen small enough so as to make $1-\text{erf}(r/a)$ 
negligible at the cell boundaries. The second term can be calculated 
analytically

\begin{equation}
\displaystyle \int d{\bf r} \frac{1-\text{erf}(r/a)}{r}e^{-i{\bf Gr}} =
\frac{4\pi}{G^2} \{1 - e^{-G^2 a^2 /4} \}
\end{equation}
and the numerical integration reduces to the first term, which is free of 
singularities. 
Even so, this term cannot be calculated by simply applying an FFT because the 
repeated function, although periodic, is not analytic at the boundary.
Use of the FFT amounts to using the trapezium rule for the integration,
which is
exact for a periodic analytic function, but leads to substantial errors when
there are discontinuous derivatives as we have in this case.
We evaluated the integral using a second-order Filon's method, \cite{Filon}
which proved to be effective. 
Other procedures \cite{numerical_recipes} (Simpson's, 
Romberg's...) could have been used -  they are all rather slow if accurate
results are to be obtained, but this calculation needs to be done only once 
for a cubic cell. If we denote by 
$I[D(L)](n_1,n_2,n_3)$ the integral in Eq.\ (\ref{eq:cubic_cutoff_int})
for a cubic box of side $L$ and frequency indices $(n_1,n_2,n_3)$, 
it is clear that
$I[D(L)](n_1,n_2,n_3) = L^2 I[D(1)](n_1,n_2,n_3)$.

\section{Results}

\subsection{Exactly soluble systems}

It is interesting to see the effect of the various multipoles that a charge 
distribution might have by using the multipoles-correction method
on an exactly soluble system. We have studied systems consisting of 
superpositions of Gaussian charge distributions placed at various points 
${\bf R}_i$ within the cubic cell of side $L$:

\begin{equation} \label{eq:gaussians}
n({\bf r})=\sum_{i} z_i\frac{\text{exp}( -\frac{|{\bf r}-{\bf R}_i|^2}
{a_{i}^{2}} )}
{a_{i}^{3} \pi^{3/2}}.
\end{equation}

\begin{figure}[h]
\centerline{\hbox{\psfig{figure=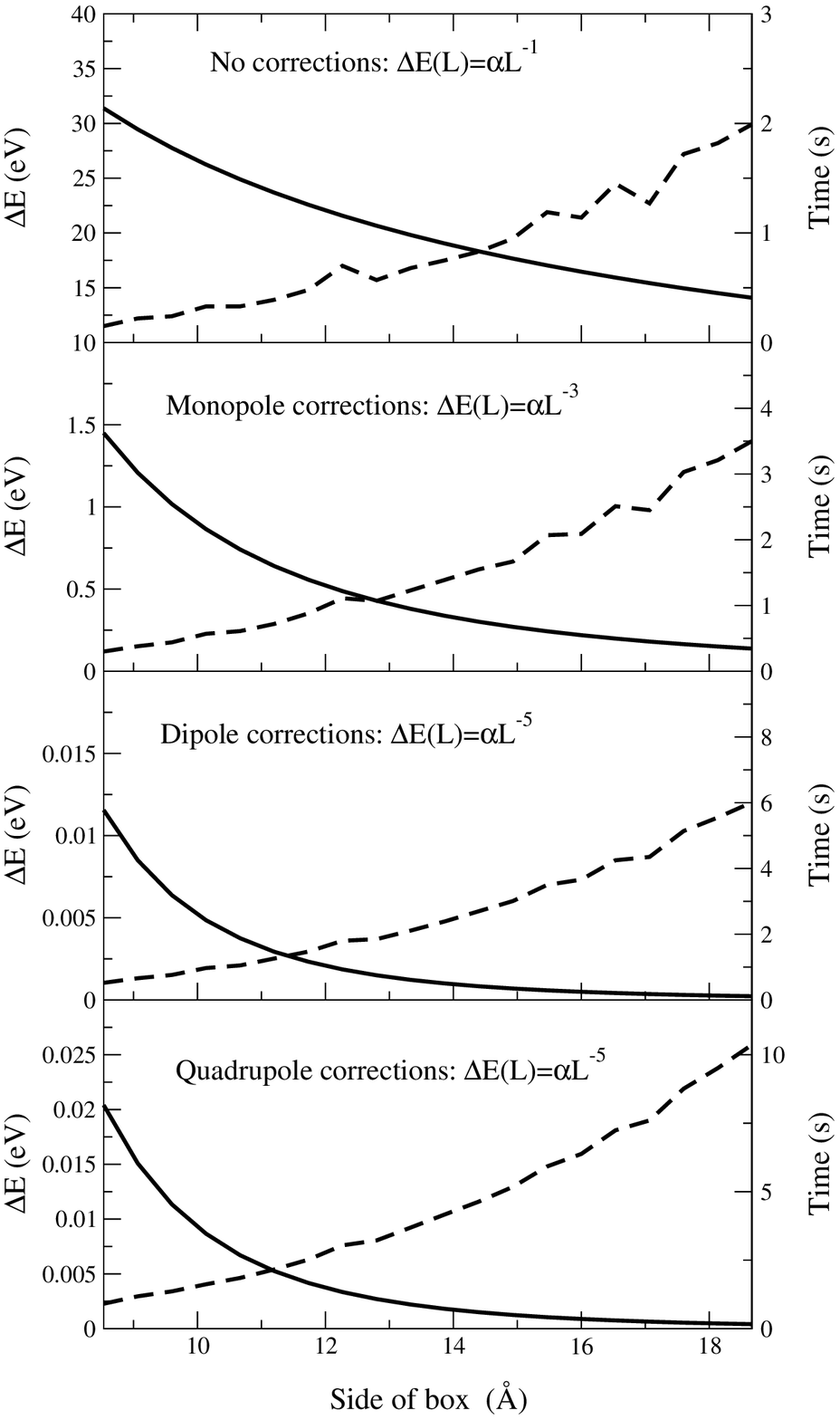,width=4in}}}
\caption{
Error in the electrostatic energy for the system of Gaussian charges,
Eq. (\ref{eq:gaussians}), (continuous line) 
and total time of
calculations (dashed line), for the indicated order used of the 
multipoles correction.}
\label{figure1}
\end{figure}

We have investigated the efficiency with which the multipole corrections 
remove the effects of the images of the charge distributions in other cells.
This has been done as functions of L, as for a large enough cell the results
for the periodic system should become exact, but at rates depending on the 
order of the multipoles. The results are shown in Fig.\ \ref{figure1}
for the cases in which there are (i) no corrections (by which we mean that only
the constant to meet the proper boundary conditions is added to the raw
potential obtained from the Fourier transform), (ii) monopole corrections, 
(iii) monopole + dipole corrections, and (iv) monopole + dipole + quadrupole 
corrections. The following points are noteworthy. 

\begin{itemize}

\item There is a serious, roughly 10\%, error in the total energy when the
Hartee potential is uncorrected. Although this is not a consideration in
superlattice calculations provided the system is neutral, it is an important
matter in real space calculations when the Hartree potentials due to the 
electrons alone is calculated in reciprocal space.

\item The time for the calculations behaves roughly 
as $O(L^{3}\text{log}L)$, but with irregularities. The efficiency of 
the FFT algorithm depends on the prime factorization of the
number of points to be transformed. The original FFT was developed for
powers of two, but now 
algorithms exist with more
flexibility.\cite{Oppenheim_Shafer_1989,Duharnel_Vetterli}
We have used the FFTW package,\cite{Frigo_Johnsons_98} with support
for all the primes involved in our calculations.

\item Adding the quadrupole corrections does not seem to improve the 
accuracy of the result, nor is the $L$-dependence improved. This is because
the interaction energy between the dipole of the charge distribution
in the central cell and octupoles in other cells, has the same
$L$-dependence and order of magnitude as the quadrupole-quadrupole energy. 
Consequently, although the potential will be improved by adding to it the 
quadrupole corrections, there could be no significant improvement in the 
total energy if the system has a strong dipole.
In general, it can be shown that the error in the electrostatic energy  
due to the presence of an $l$ multipole in the charge distribution in the
central cell, and $l'$ multipoles in all other cells goes to zero like 
$L^{-(l+l'+1)}$, or in some special cases faster due to symmetry (for 
instance, if $l=0$ and $l'\leq3$, or $l$ and $l'$ have different parity). 
Thus, adding octupole corrections to the potential will not change the 
$L$-dependence of the total energy if the system is charged  because of the 
interaction of the monopole with the $l=4$ multipole. 
Our calculations below on the
$\text{Na}_{10}^{+2}$ cluster provide an interesting example of this behaviour.
\end{itemize}

\subsection{Real systems}

We have performed several electronic structure calculations on real systems
to assess the performance of the methods. We have used a real-space
code,\cite{Bertsch_etal_00}
in which a superlattice and plane waves are only used to accelerate
the solution of Poisson's equation for the electron charge distribution. In 
this type of approach  a correction for the net charge is always needed 
irregardless of whether the molecule or cluster itself is charged or neutral. 
Furthermore, in this approach the value of the multipoles will depend on the
position of the molecule with respect to the centre of the cell. In order 
to minimize the multipole corrections the centre of charge of the 
system of ions should be placed at the centre of the cell. If this is not 
done in real space calculations the errors caused by cell-to-cell 
interactions could be magnified. In order to illustrate the effects we take 
the center of charge as the cell center for one of our test cases, and not 
for the other. As for other details of the calculations we used 
density-functional theory with the local-density approximation for exchange and
correlation, and Troullier-Martins\cite{Troullier_Martins_91} 
nonlocal, norm-conserving pseudopotentials.

Our first choice for a realistic system was the NaCl molecule, also treated
by Shultz \cite{Shultz_99} 
and Jarvis {\em et al},\cite{Jarvis_etal_97} 
because of its strong dipole moment (experimental value of 
9.0D in the gas phase, as reported by Nelson {\em et al}.\cite{Nelson})
In this case, the center of charge of the system of ions is placed at the center of 
the cell. The equilibrium bond-length was calculated: (i) using the spherical 
cut-off method which is exact with a large enough cutoff, and (ii) using the
multipoles correction and correcting only for the monopole term so as to
show the influence of the dipole-dipole interactions which are ignored.
Our calculated ``exact'' value is 2.413\AA, whereas the result ignoring the 
dipole-dipole interactions is 2.448\AA.

Next, we investigated the performance and accuracy of the methods by 
determining the errors in the total energy and electric dipole moment, while
monitoring the calculation times. We compared results for the energy and
dipole moment against those obtained with the spherical cutoff method with 
a cut-off radius of $\sqrt{3}L_C$, grid parameter (0.2\AA) and cell size 
($L$=10\AA). In this way an electric dipole moment of 8.4551D was obtained.
Each of the four methods was then used to converge the electronic ground 
state of the molecule for successive values of a ``control parameter'' for 
speed and accuracy:

\begin{itemize}

\item For the real-space, conjugate-gradients method this parameter was the 
order of the difference formula used to evaluate derivatives. 

\item For the spherical cut-off method, we note that, if the electron
density is well localized within the $C$ cell, the need for the full cut-off 
radius, $\sqrt{3}L_C$, may be relaxed and a correspondingly 
smaller $D$ cell used, introducing some error but accrueing time savings. 
We have investigated the effect of using a reduced cut-off radius, 
$r_{\text{cut-off}}$, through a control parameter, $\alpha$, which is the 
ratio of the $D$ and $C$ cubic cell edges: 

\begin{equation} 
\alpha=\frac{L_D}{L_C}=1+\frac{r_{\text{cut-off}}}{L_C}
\end{equation}

so that
$\alpha=1+\sqrt{3}$ is the minimum value for which exact results are 
guaranteed. 

\item The $D$ cell size can also be reduced in the case of the cubic cut-off 
method, and the control parameter is again $ \alpha=\frac{L_D}{L_C}$
where $\alpha\ge 2$ guarantees exact results.

\item For the multipoles correction method, the order of the multipoles 
corrected for is the control parameter.

\end{itemize}

\begin{figure}[h]
\centerline{\hbox{\psfig{figure=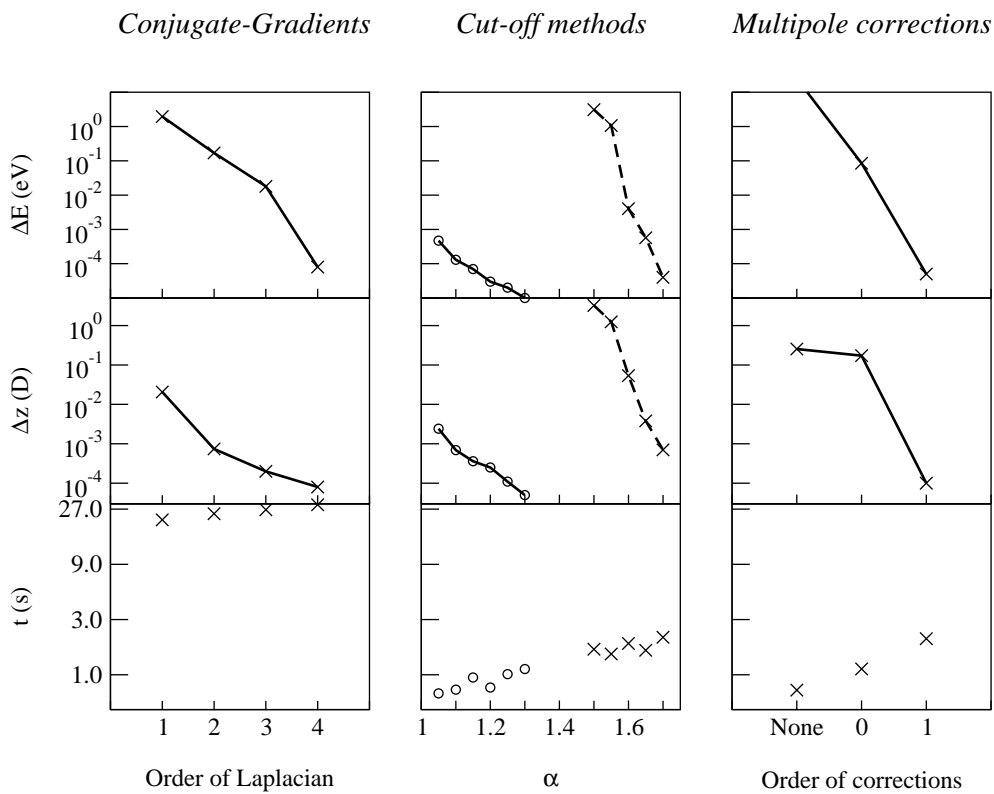,width=6in,angle=-90}}}
\caption{
Error in the electrostatic (first row), electric dipole
(second row) and time of calculations (third row) for the 
NaCl molecule, using the 
methods indicated, as a function of the respective ``control parameter'' 
(see text). For the cut-off methods, crosses refer to the spherical 
cut-off method, and circles to the cubic cut-off method. All scales are
logarithmic.}
\label{figure2}
\end{figure}

In Fig.\ \ref{figure2} we illustrate the results obtained for each of the 
methods. Both cut-off methods are presented in the same column as they 
use the same control parameter, although the ranges of values are different.

\begin{enumerate}

\item The real-space method is significantly slower than the other
methods for the same accuracy, and a case can be made for using reciprocal
space methods for calculating the Hartree potential in what are otherwise 
real-space codes. However, enhancements of the conjugate-gradients method
are possible through preconditioning and multigrid 
techniques.\cite{Payne_etal_?,Beck_2000}

\item The cut-off methods reach acceptable accuracy much below
the values for $\alpha$ which guarantee exact results: $1+\sqrt{3}$ and $2$ 
respectively, for the spherical and cubic cut-off. This to be expected when the
charge distribution is well localized within the cell.
However, it is clearly demonstrated that, for a given accuracy, the size 
of the auxiliary cell is smaller for the cubic cut-off method,
and as a result, the calculation time is also shorter.

\item The multipoles-correction method already gives good accuracy 
if the dipole interactions are corrected for ($5\times 10^{-5}$eV error in 
the energy, and $10^{-4}$D error in the electric dipole). Without the 
dipole correction, the error in the energy is $0.085$eV, and in the
dipole is 0.17D, which give an indication of the size of errors to be expected 
when supercell calculations are performed for neutral molecules and no 
corrections are made. 

\end{enumerate}

We have also performed calculations on the $\text{Na}_{10}^{+2}$ cluster 
containing the same number of valence electrons as the NaCl molecule. 
Results are similar to those presented for NaCl, but some differences 
should be reported. In this case the center of charge was not placed 
at the center of the cell, consequently, although the cluster has a calculated 
net dipole of $4.5D$, the electronic dipole responsible for the errors
is a much larger $10.2D$. The cluster was positioned in the cell so that the 
charge density occupied most of the cell, allowing an optimally small cell.
As a result, to achieve similar accuracy as for the NaCl molecule, we 
should expect the need for (i) larger cut-off lengths for the cut-off methods, 
and (ii) higher multipole corrections.

\begin{figure}
\centerline{\hbox{\psfig{figure=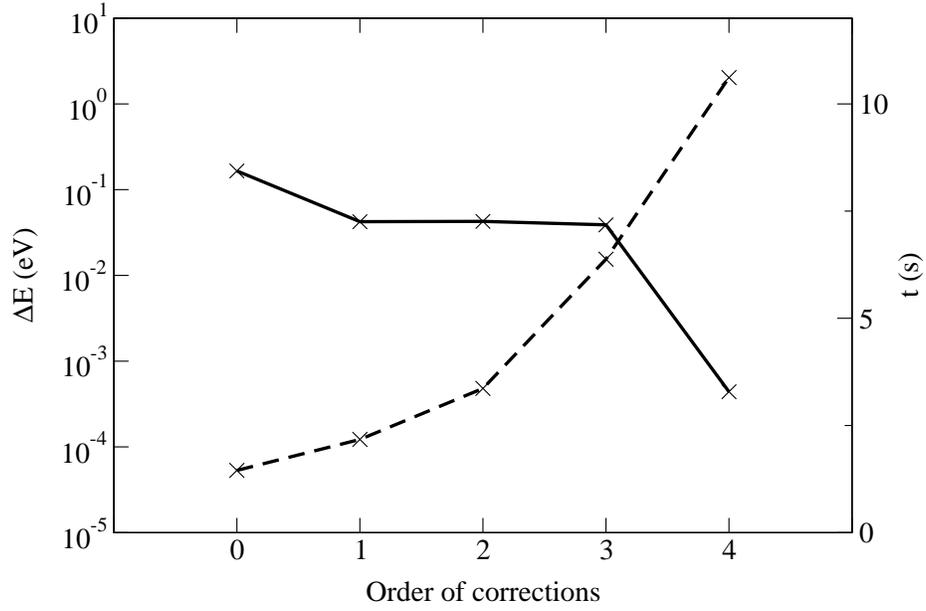,width=5in,angle=-90}}}
\caption{Error in the electrostatic energy (continuous line) 
for the $\text{Na}_{10}^{+2}$
cluster,
using the multipoles correction method, as a function of the order
of corrections included in the calculations. Also shown is the time of 
calculation for each case (dashed line). Note that the time scale is 
not logarithmic.}
\label{figure3}
\end{figure}

In Fig.\ \ref{figure3} we show the error in the energy obtained by 
using the multipoles correction method. It can be seen how the inclusion of
the dipole correction yields a much less satisfactory error in the energy
than for NaCl. Furthermore, for the reasons given earlier, the removal 
of the quadrupole-quadrupole, dipole-octupole, and octupole-octupole terms
does not significantly improve the accuracy. Only by including all 
corrections to the potential up to fourth order multipole moments do we 
obtain a comparable result for the energy. The calculation time, which is 
also shown in the figure, is beginning to increase sharply by fourth order as 
further corrections are added.

\begin{figure}[h]
\centerline{\hbox{\psfig{figure=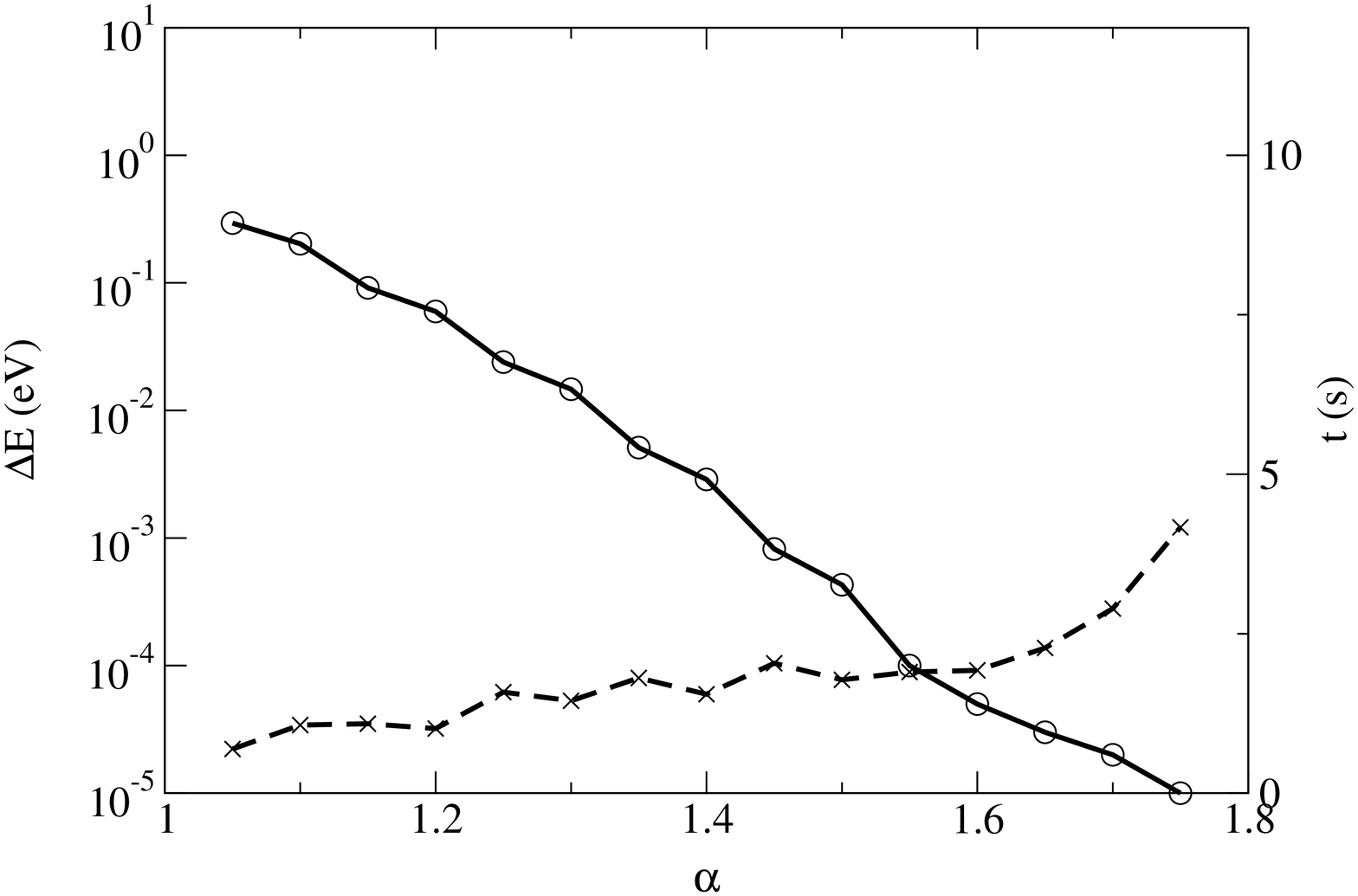,width=5in,angle=-90}}}
\caption{Error in the electrostatic energy for the $\text{Na}_{10}^{+2}$ 
cluster
using the cubic cut-off method, as a function of its ``control
parameter''. Also shown is the time of calculation for each case (dashed line).
Note that the time scale is not logarithmic.}
\label{figure4}
\end{figure}

In Fig.\ \ref{figure4} we present, as well, the results for the error in the 
total energy and the calculation time for the cubic-cutoff method. Comparison
with Fig.\ \ref{figure2} confirms that the energy converges 
much less rapidly as a function
of $\alpha$ than for the NaCl molecule.

\section{Conclusions.}

We have studied some of the methods which have been proposed recently for
solving Poisson's equation in reciprocal space  
for electronic structure calculations on finite systems. We also propose a 
method based on multipole corrections. 
Test calculations have been performed to assess the performance of the methods.
We conclude that reciprocal-space methods can be accurate
enough for finite or aperiodic systems, and their efficiency is a significant 
improvement over that of real-space methods. Two basic reciprocal-space
methods have been investigated: one which imposes a cut-off on 
the Coulomb potential, and one based on the removal of the spurious
effects through a multipole expansion. Both yield satisfactory results, and 
comparable efficiency.

The former approach has been already been surveyed by Jarvis {\em et al}
\hspace{0.1cm}. \cite{Jarvis_etal_97} There are two possibilities for the
cut-off function, one, the spherical cut-off, was highlighted for having 
superior convergence with the plane wave cut-off of the reciprocal 
lattice. However, we have identified and corrected problems with 
the other possibility, the cubic cut-off, which eliminates the poor behaviour, 
and makes this cut-off method the better of the two because smaller FFT's are
allowed. 

The method based on multipoles corrections was initiated by Shultz, 
\cite{Shultz_99} but we have developed a scheme which we think is more 
general. Its performance is more predictable than that of the cut-off methods, 
which are sensitive to the choice of the cut-off length. The reason
for the sensitivity is that the cut-off length determines the size of a larger
auxiliary cell and the number of grid points over which FFT calculations are 
performed, and the FFT is sensitive to the prime number decomposition
of the number of points. On the other hand, the speed and accuracy of our 
``multipoles correction'' method, are adequate for most applications, and can 
be easily controlled by choosing the order of corrections applied.

All the methods have been presented assuming a cubic cell. However, 
generalizations to other cell shapes are possible if the geometry of the 
system requires it. The multipoles correction method is immediately 
applicable to any cell. Clearly the spherical cut-off method would be
inefficient for elongated cells because the radius of the cut-off sphere
is determined by the longest dimension of the cell. But, the cubic cut-off 
method can easily be generalized to other cell shapes, at the cost of
more, and more lengthy calculations of the Fourier transforms of the 
truncated Coulomb interaction.

We have made a simple implementation of the solvers within the 
self-consistent framework, but smarter algorithms can be developed,
since not all the iterations of a self-consistent calculation need be
done with the same accuracy. For example, significant improvements in 
efficiency can be gained if, for a given method, the iterations are
started with a fast but inexact solver through appropriate choice
of the control parameter, but improved as self-consistency is
approached. Moreover, methods could be combined using, for instance,
the real-space method for the last few iterations because of its
efficiency when a good starting point is known.

\acknowledgments

We are pleased to acknowledge useful discussions with J. A. Alonso.
We also acknowledge financial support from 
JCyL (Grant: VA 28/99) and DGES (Grant No. DGES-PB98-0345) and 
European Community TMR contract NANOPHASE: HPRN-CT-2000-00167.
A. C. acknowledges 
financial support by the MEC, and 
hospitality by Queen's University, where 
most of this work has been done, during a research visit.
M. J. S. acknowledges the support of the NSERC of Canada, of 
Iberdrola through its visiting professor program, and of 
the Universidad de Valladolid where this work began.

\begin{references}


\bibitem{Hohenberg_Kohn_64} P. Hohenberg and W. Kohn, Phys. Rev. {\bf 136},
B864 (1964).

\bibitem{Kohn_Sham_65} W. Kohn and L. J. Sham, Phys. Rev. {\bf 140},
A1133 (1965).

\bibitem{Runge_Gross_84} E. Runge and E. K. U. Gross, Phys. Rev. Lett.
{\bf 52}, 997 (1984).

\bibitem{Rubio_etal_97} A. Rubio, J. A. Alonso, X. Blase and S. G. Louie, 
Int. J. Mod. Phys. B {\bf 11}, 2727 (1997).

\bibitem{Brack_93} M. Brack, Rev. Mod. Phys. {\bf 65}, 677 (1993).

\bibitem{Pickett_89} W. E. Pickett, Comput. Phys. Rep. {\bf 9}, 115 (1989).

\bibitem{Cohen_82} M. L. Cohen, Solid. State Commun. {\bf 92}, 45 (1994);
Phys. Scri. {\bf 1}, 5 (1982).

\bibitem{Lauritsch_Reinhard_94} G. Lauritsch and P.-G. Reinhard, Int. J. 
Mod. Phys. C {\bf 5}, 65 (1994).

\bibitem{Onida_etal_95} G. Onida {\em et al}, Phys. Rev. Lett. {\bf 75},
818 (1995).

\bibitem{Makov_Payne_95} G. Makov and M. C. Payne, Phys. Rev. B {\bf 51},
4014 (1995).

\bibitem{Jarvis_etal_97} M. R. Jarvis, I. D. White, R. W. Goodby and 
M. C. Payne,
Phys. Rev. B {\bf 56}, 14972 (1997).

\bibitem{Shultz_99} P. A. Shultz, Phys. Rev. B {\bf 60}, 1551 (1999).  

\bibitem{Shultz_00} P. A. Shultz, Phys. Rev. Lett. {\bf 84}, 1942 (2000).

\bibitem{webpage} http://www.fam.cie.uva.es/\~{}arubio.

\bibitem{Hummer_96} G. Hummer, J. Electrostatics {\bf 36}, 285 (1996).

\bibitem{note_on_sa_1} Although a cubic cell is used for simplicity,
extension to a rectangular parallelpiped is possible. This could
be important for molecular dynamics simulation with variable cell
shape.

\bibitem{note_on_sa_2} In fact, some numerical difficulties 
arise when calculating this constant.
A better approach is splitting the function into periodic and aperiodic
parts, and performing the integration only for the periodic part, which
should integrate to zero.

\bibitem{numerical_recipes} W. H. Press, S. A. Teukolsky, W. T Vetterling
and B. P. Flannery, {\em Numerical Recipes} (Cambridge University Press,
New York).

\bibitem{Filon} L. N. G. Filon, Proceedings of the Royal Society of 
Edinburgh {\bf 49}, 38 (1928).

\bibitem{Oppenheim_Shafer_1989} A. V. Oppenheim and R. W. Schafer, 
{\em Discrete Signal Processing} (Prentice-Hall, Englewood Gliffs NV,1989).

\bibitem{Duharnel_Vetterli} P. Duharnel and M. V\'{e}tterli, 
Signal Processing {\bf 19}, 259 (1990).

\bibitem{Frigo_Johnsons_98} M. Frigo and S. G. Johnsons, in {\em Proceedings of
the IEEE International Conference on Acoustic Speech and Signal 
Processing, Seattle, Washington, 1998}, vol. 3, p. 1381.

\bibitem{Troullier_Martins_91} N. Troullier and J. L. Martins, Phys. Rev. B
{\bf 43}, 1993 (1991).

\bibitem{Bertsch_etal_00} G. F. Bertsch, J.-I. Iwata, A. Rubio and K. Yabana, Phys. Rev. B {\bf 62}, 7998 (2000). 

\bibitem{Nelson} R. D. Nelson, D. R. Lide, A. A. Maryott,
{\em National Reference Data Series - National Bureau of Standards} 
(NRDS-NBS 10).

\bibitem{Payne_etal_?} M. C. Payne, M. P. Teter, D. C. Allan, T. A. Arias
and J. D. Joannopoulos, Rev. Mod. Phys. {\bf 64}, 1045 (1992).

\bibitem{Beck_2000} T. L. Beck, Rev. Mod. Phys. {\bf 72}, 1041 (2000).





 


\end{references}
\end{document}